\documentstyle[11pt,twoside,epsfig,rotate]{article}

\newlength{\dinwidth}
\newlength{\dinmargin}
\def\psip{\psi(2S)}
\def\Jpsi{J/\psi}
\def\Wgp{W_{\gamma p}}

\begin{document}  
\newcommand{\pom}{{I\!\!P}}
\newcommand{\slowpi}{\pi_{\mathit{slow}}}
\newcommand{\gevsq}{\mathrm{GeV}^2}
\newcommand{\fiidiii}{F_2^{D(3)}}
\newcommand{\fiidiiiarg}{\fiidiii\,(\beta,\,Q^2,\,x)}
\newcommand{\n}{1.19\pm 0.06 (stat.) \pm0.07 (syst.)}
\newcommand{\nz}{1.30\pm 0.08 (stat.)^{+0.08}_{-0.14} (syst.)}
\newcommand{\fiidiiiful}{F_2^{D(4)}\,(\beta,\,Q^2,\,x,\,t)}
\newcommand{\fiipom}{\tilde F_2^D}
\newcommand{\ALPHA}{1.10\pm0.03 (stat.) \pm0.04 (syst.)}
\newcommand{\ALPHAZ}{1.15\pm0.04 (stat.)^{+0.04}_{-0.07} (syst.)}
\newcommand{\fiipomarg}{\fiipom\,(\beta,\,Q^2)}
\newcommand{\pomflux}{f_{\pom / p}}
\newcommand{\nxpom}{1.19\pm 0.06 (stat.) \pm0.07 (syst.)}
\newcommand {\gapprox}
   {\raisebox{-0.7ex}{$\stackrel {\textstyle>}{\sim}$}}
\newcommand {\lapprox}
   {\raisebox{-0.7ex}{$\stackrel {\textstyle<}{\sim}$}}
\newcommand{\pomfluxarg}{f_{\pom / p}\,(x_\pom)}
\newcommand{\dsf}{\mbox{$F_2^{D(3)}$}}
\newcommand{\dsfva}{\mbox{$F_2^{D(3)}(\beta,Q^2,x_{I\!\!P})$}}
\newcommand{\dsfvb}{\mbox{$F_2^{D(3)}(\beta,Q^2,x)$}}
\newcommand{\dsfpom}{$F_2^{I\!\!P}$}
\newcommand{\gap}{\stackrel{>}{\sim}}
\newcommand{\lap}{\stackrel{<}{\sim}}
\newcommand{\fem}{$F_2^{em}$}
\newcommand{\tsnmp}{$\tilde{\sigma}_{NC}(e^{\mp})$}
\newcommand{\tsnm}{$\tilde{\sigma}_{NC}(e^-)$}
\newcommand{\tsnp}{$\tilde{\sigma}_{NC}(e^+)$}
\newcommand{\st}{$\star$}
\newcommand{\sst}{$\star \star$}
\newcommand{\ssst}{$\star \star \star$}
\newcommand{\sssst}{$\star \star \star \star$}
\newcommand{\tw}{\theta_W}
\newcommand{\sw}{\sin{\theta_W}}
\newcommand{\cw}{\cos{\theta_W}}
\newcommand{\sww}{\sin^2{\theta_W}}
\newcommand{\cww}{\cos^2{\theta_W}}
\newcommand{\trm}{m_{\perp}}
\newcommand{\trp}{p_{\perp}}
\newcommand{\trmm}{m_{\perp}^2}
\newcommand{\trpp}{p_{\perp}^2}
\newcommand{\alp}{\alpha}
\newcommand{\alps}{\alpha_s}
\newcommand{\sqrts}{$\sqrt{s}$}
\newcommand{\LO}{$O(\alpha_s^0)$}
\newcommand{\Oa}{$O(\alpha_s)$}
\newcommand{\Oaa}{$O(\alpha_s^2)$}
\newcommand{\PT}{p_{\perp}}
\newcommand{\JPSI}{J/\psi}
\newcommand{\GP}{\gamma p  \rightarrow }
\newcommand{\Gg}{\gamma g  \rightarrow }
\newcommand{\sh}{\hat{s}}
\newcommand{\uh}{\hat{u}}
\newcommand{\MP}{m_{J/\psi}}
\newcommand{\PO}{I\!\!P}
\newcommand{\xbj}{x}
\newcommand{\xpom}{x_{\PO}}
\newcommand{\ttbs}{\char'134}
\newcommand{\AmS}{{\protect\the\textfont2
  A\kern-.1667em\lower.5ex\hbox{M}\kern-.125emS}}
\newcommand{\xpomlo}{3\times10^{-4}}  
\newcommand{\xpomup}{0.05}  

\begin{titlepage}
\begin{flushleft}
{\tt DESY 97-228}\hfill {\tt ISSN 0418-9833} \\
\end{flushleft}
\vspace*{3.0cm}
\begin{center}\begin{LARGE}
\boldmath{\bf Photo-production of $\psip$ Mesons at HERA}\unboldmath \\
\vspace*{2.5cm}
H1 Collaboration \\
\vspace*{2.5cm}
\end{LARGE}
{\bf Abstract}
\begin{quotation}
\noindent
Quasi-elastic ($z >0.95$)
photo-production of $\psip$ mesons has been observed at HERA 
for  
photon-proton centre-of-mass energies in the range $40$ to $160$~GeV. 
The $\psip$
mesons were identified through their decays to 
$\ell^{+} \ell^{-}$ and to $\Jpsi \, \pi^{+} \pi^{-}$, where the $\Jpsi$
subsequently decays to $\ell^{+} \ell^{-}$, the lepton $\ell$ being either 
a muon or an electron. The cross-section for quasi-elastic photoproduction
was measured to be $(18.0\pm 2.8({\rm stat}) \pm 3.0({\rm syst}))$~nb at a 
photon-proton centre-of-mass energy of 80 GeV. The ratio of 
the $\psip$ to $\Jpsi$ quasi-elastic
cross-sections is $(0.150 \pm 0.027({\rm stat}) \pm 0.022({\rm syst}))$. 
\end{quotation}
\vspace*{2.0cm}
{\it Submitted to Physics Letters $\mbox{\boldmath $B$}$}  \\
\vfill
\cleardoublepage
\end{center}
\end{titlepage}

\vfill
\clearpage
\begin{sloppypar}
\noindent
 C.~Adloff$^{35}$,                
 S.~Aid$^{13}$,                   
 M.~Anderson$^{23}$,              
 V.~Andreev$^{26}$,               
 B.~Andrieu$^{29}$,               
 V.~Arkadov$^{36}$,               
 C.~Arndt$^{11}$,                 
 I.~Ayyaz$^{30}$,                 
 A.~Babaev$^{25}$,                
 J.~B\"ahr$^{36}$
% ...
 \\
\bigskip 
 
\noindent
{\footnotesize{
% ...
$ ^7$ Physics Department and IIRPA,
       University of California, Davis, California, USA$^e$ \\
% ...
$ ^{11}$ DESY, Hamburg, Germany$^a$ \\
% ...
$ ^{13}$ II. Institut f\"ur Experimentalphysik, Universit\"at Hamburg,
          Hamburg, Germany$^a$  \\
% ...
$ ^{23}$ Physics Department, University of Manchester, Manchester, UK$^b$ \\
% ...
$ ^{25}$ Institute for Theoretical and Experimental Physics,
          Moscow, Russia \\
 $ ^{26}$ Lebedev Physical Institute, Moscow, Russia$^{f,k}$ \\
% ...
$ ^{29}$ LPNHE, Ecole Polytechnique, IN2P3-CNRS, Palaiseau, France \\
 $ ^{30}$ LPNHE, Universit\'{e}s Paris VI and VII, IN2P3-CNRS,
          Paris, France \\
% ...
$ ^{35}$ Fachbereich Physik, Bergische Universit\"at Gesamthochschule
          Wuppertal, Wuppertal, Germany$^a$ \\
 $ ^{36}$ DESY, Institut f\"ur Hochenergiephysik, Zeuthen, Germany$^a$ \\
% ...
$ ^a$ Supported by the Bundesministerium f\"ur Bildung, Wissenschaft,
        Forschung und Technologie, FRG,
        under contract numbers 7AC17P, 7AC47P, 7DO55P, 7HH17I, 7HH27P,
        7HD17P, 7HD27P, 7KI17I, 6MP17I and 7WT87P \\
 $ ^b$ Supported by the UK Particle Physics and Astronomy Research
       Council, and formerly by the UK Science and Engineering Research
       Council \\
% ...
$ ^e$ Supported in part by US~DOE grant DE~F603~91ER40674 \\
 $ ^f$ Supported by the Deutsche Forschungsgemeinschaft \\
% ...
$ ^k$ Supported by Russian Foundation for Basic Researches 
       grant no. 96-02-00019 \\
}}
 
\end{sloppypar}
 
 
\newpage
 

\section{Introduction} \label{sect:intro}

At HERA the production of vector mesons is studied at 
photon-proton centre-of-mass
energies ($W_{\gamma p}$) extending well beyond those reached in fixed 
target experiments. 
In particular, the production of $\Jpsi$ mesons has been
reported upon by H1 \cite{ref:H1_Jpsi1,ref:H1_Jpsi2,
ref:H1_Jpsi3} and ZEUS \cite{ref:Zeus_Jpsi1,ref:Zeus_Jpsi2,ref:Zeus_Jpsi3}.
In this letter a study of the quasi-elastic photo-production of
$\psip$, i.e. $\psi$(3685), mesons in the photon-proton centre-of-mass 
energy range
$40<\Wgp<160$~GeV is reported. 
Quasi-elastic is defined here as the kinematic region $z>0.95$, where
$z = E_{\psip}/E_{\gamma}$ in the rest frame of the proton.
The $\psip$ mesons are identified
via the decays $\psip \rightarrow J/\psi (\rightarrow \ell^{+}\ell^{-})
\pi ^{+} \pi^{-}$ or $\psip \rightarrow \ell^{+}\ell^{-}$, where $\ell$ denotes
a muon or an electron.

\begin{figure}[h]
\setlength{\unitlength}{1cm}
\begin{center}
\begin{picture}(8.0,2.8)
\put(0.0,0.0){\epsfig{figure=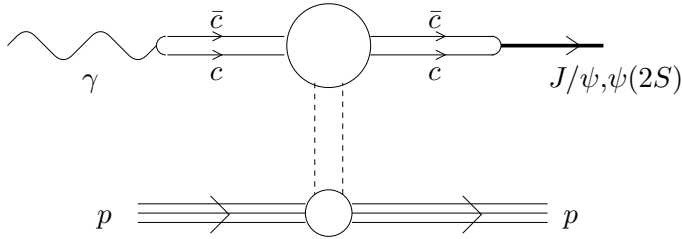,width=8.0cm}}
\put(1.0,2.0){$\gamma$}
\put(1.2,0.2){$p$}
\put(7.4,0.2){$p$}
\put(2.7,2.1){$c$}
\put(2.7,2.8){$\bar{c}$}
\put(5.6,2.1){$c$}
\put(5.6,2.8){$\bar{c}$}
\put(7.2,2.0){$\Jpsi$,$\psip$}
\end{picture}
\end{center}
\caption{Elastic production of charmonium}
\label{fig:vprodn}
\end{figure}

At the collision energies studied here, the process $\gamma p \rightarrow Vp$
($V$ = $\Jpsi$ or $\psip$) can be considered in three
parts~(see figure~\ref{fig:vprodn}): the formation of a 
$c \bar{c}$ quark pair from the photon,
the interaction of the quark pair with the proton, and
subsequently the formation of the charmonium state. 
In a QCD based model~\cite{ref:Kop91,ref:Kop93} of charmonium production
described in section~\ref{sect:cdcalc}, 
this is justified by arguing that, viewed in the proton rest frame, the
$c \bar{c}$ quark pair creation from the photon
and the formation of the charmonium state from the $c \bar{c}$ pair both
take place over distances
considerably greater than the proton radius. The $c \bar{c}$ quark pair is
regarded as a colour dipole which interacts with the proton. A prediction of
the ratio of the cross-sections for
$\psip$ and $\Jpsi$ mesons arising from this model is compared to
the measured ratio, presented in section~\ref{sect:ratio}.


\section{Kinematics} \label{sect:kine}

\begin{figure}[t]
\setlength{\unitlength}{1cm}
\begin{center}
\begin{picture}(8.0,3.0)
\put(-3.0,-0.7){\epsfig{figure=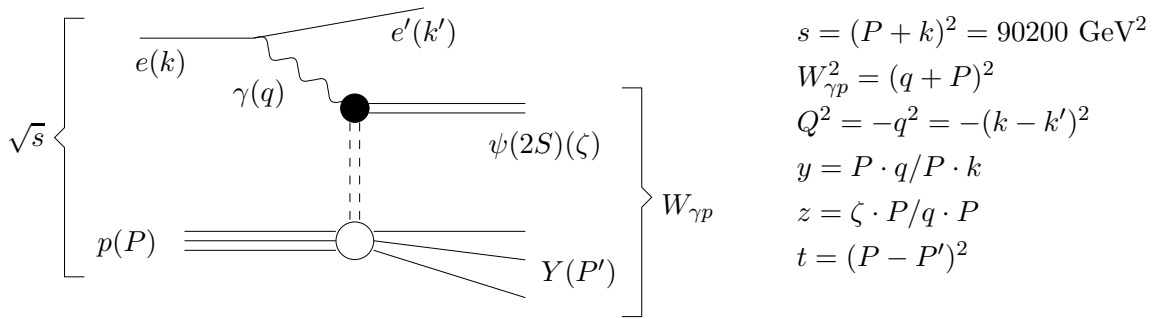,width=8.0cm}}
\put(-3.6,1.6){$\sqrt{s}$}
\put(-1.9,2.6){$e(k)$}
\put(1.5,3.0){$e^{\prime}(k^\prime)$}
\put(-0.6,2,2){$\gamma (q)$}
\put(2.8,1.5){$\psip (\zeta)$}
\put(-2.4,0.2){$p(P)$}
\put(3.5,-0.2){$Y (P^{\prime})$}
\put(5.1,0.7){$W_{\gamma p}$}
\put(6.9,3.0){$s=(P+k)^{2} = 90200$~GeV$^2$}
\put(6.9,2.4){$W_{\gamma p}^{2} = (q+P)^{2}$}
\put(6.9,1.8){$Q^{2} = -q^{2} = -(k-k^{\prime})^{2}$}
\put(6.9,1.2){$y=P \cdot q / P \cdot k$}
\put(6.9,0.6){$z= \zeta \cdot P / q \cdot P$}
\put(6.9,0.0){$t=(P-P^{\prime})^{2}$}
\end{picture}
\end{center}
\caption{The kinematic variables describing $\psip$ production at HERA.
The incoming positron and proton are denoted $e$ and $p$ respectively,
the exchanged photon by $\gamma$, the outgoing positron by $e^{\prime}$, 
and the $\psip$ meson produced by $\psip$. The remainder of the final
state is denoted $Y$.
The associated four-momenta are indicated in brackets in the diagram.}
\label{fig:kine}
\end{figure}

The relevant kinematic quantities are defined in figure~\ref{fig:kine}.
For this study, it is demanded that neither the system $Y$ nor the
scattered positron $e^{\prime}$ are observed (see section~\ref{sect:ident}),
escaping in the $+z$~(incoming proton) and $-z$~(incoming positron)
directions respectively. The system $Y$ is the scattered proton in
the elastic case, and otherwise can be considered as the dissociated proton.

The photon-proton centre-of-mass energy, $W_{\gamma p}$, is determined using
the relation:
\begin{equation}
 W_{\gamma p}^{2} \approx \frac{s \sum (E-p_{z})}{2E_{e}}
\end{equation}
where $E_e$ is the incoming positron energy, and the sum is over the
energies, $E$, and the longitudinal momenta, $p_z$, of the decay products 
of the $\psip$ meson. The centre-of-mass energy of the positron-proton
collision is denoted $s$. Neither the elasticity of
the interaction, $z$, nor the photon virtuality, $Q^2$, can be
measured. However, the selection criteria restrict the acceptance (see 
section~\ref{sect:MC})
approximately to the kinematic region $z>0.95$, called here quasi-elastic,
 and $Q^2 < 4$~GeV$^2$; the median $Q^2$ is expected to be approximately
$10^{-4}$~GeV$^2$. Cross-sections are evaluated and presented 
in the photo-production
limit, $Q^2 = 0$~GeV$^2$, in section~\ref{sect:Xsects}.
All cross-sections are integrated over the
momentum transfer variable $t$.


\section{Experimental Conditions}
The data presented here were collected in 1994 and 1995 with the 
H1 detector at HERA, when 27.5 GeV positrons were 
in collision with 820 GeV protons. They correspond to
an integrated luminosity of 2.7 pb$^{-1}$ in 1994 and 3.6 pb$^{-1}$
in 1995.  

The H1 detector is described in detail elsewhere \cite{ref:H1_det} 
and only components relevant to the present analysis are described here.
Polar angles are defined with respect to the direction of the
incoming proton beam.
The Central Tracker (CT) surrounds the interaction region, covering
the polar angle range 
20$^{\circ}$ to 165$^{\circ}$. Multiwire proportional chambers (CIP
and COP) provide additional information for triggering in this region.
Forward of the CT is the Forward Tracker (FT) covering 
the polar angular range 7$^{\circ}$ to 25$^{\circ}$. 
Surrounding the trackers
is a highly segmented liquid argon (LAr) sampling calorimeter
\cite{ref:H1_Lar} consisting of an inner electromagnetic section
and an outer hadronic section. 
Polar angle coverage is from 4$^{\circ}$ to 153$^{\circ}$. 
During the 1994 data taking a lead scintillator 
electromagnetic calorimeter (BEMC)
covered the angular range 155$^{\circ}$ to 176$^{\circ}$ 
and was able to detect
scattered positrons down to a $Q^{2}$ of 4 GeV$^{2}$
\cite{ref:BEMC}. This calorimeter was replaced
for 1995 data taking by a scintillating fibre and lead calorimeter (SpaCal) 
with coverage of the angular range 153$^{\circ}$ to 177.5$^{\circ}$, thus
extending the acceptance down to a $Q^{2}$ of 1.2 GeV$^{2}$
\cite{ref:H1_spacal}. 
The tracking chambers are immersed in a uniform axial magnetic field of
1.15~T provided by a superconducting magnet that 
surrounds the LAr calorimeter. Outside this lies the iron 
return yoke of the magnet which is instrumented to detect muons in
the range 4$^{\circ}$ to 171$^{\circ}$. 
Luminosity is determined by measuring the rate of the reaction
$ep \rightarrow ep \gamma$.

A first level trigger accepts events in which a $\psip$ meson
is produced if a track of at least $600$~MeV momentum pointing 
towards an energy cluster greater than about $1$~GeV
in the LAr calorimeter is identified, or two tracks back-to-back
in azimuth in the multiwire proportional chambers (CIP and COP)
are found, or if there exists a  muon signature in the instrumented iron 
of the return yoke. 


\section{Identification of the \boldmath{$\psip$}} \label{sect:ident}

The $\psip$ meson is identified by its decays to
$\Jpsi (\rightarrow e^{+}~e^{-})\pi^{+}~\pi^{-}$, 
$\Jpsi (\rightarrow \mu^{+}~\mu^{-})\pi^{+}~\pi^{-}$,
$e^{+}~e^{-}$ and $\mu^{+}~\mu^{-}$ which have measured branching ratios of
1.95$\pm$0.17\%, 1.95$\pm$0.17\%, 0.88$\pm$0.13\%
and 0.77$\pm$0.17\% respectively \cite{ref:PDG}.
 
The decay products of the $\psip$ are measured in the CT and are
required to
have $p_t > 120$~MeV and $20^{\circ} < \theta <  165^{\circ}$.
Particle identification is provided for the leptons by the LAr
calorimeter, in which energy deposits typical for muons or
electrons are observed, and for muons by tracks found in
the instrumented iron return yoke. It is demanded that the leptons 
have measured transverse momenta
greater than $800$~MeV. In the case of the four-particle decays of
the $\psip$, the lepton pair must have a measured effective mass
between $2.8$ and $3.4$~GeV. No particle identification
is required for the pions.
It is required that no charged particles other than the decay
products of the $\psip$ are observed in the CT or FT. 
A measured photon-proton centre-of-mass
energy, $\Wgp$, in the range $40-160$ ($40-120$)~GeV for the
dimuon (dielectron) sample is demanded (see section~\ref{sect:MC}).
 
Events in which the scattered positron is observed in the backward 
electromagnetic
calorimeter (BEMC in 1994, SpaCal in 1995) or in the LAr calorimeter,
are rejected, which
restricts the acceptance to the kinematic range  
$Q^{2}$ $<$ 4 (1.2) GeV$^{2}$ for the 1994 (1995) data sample. Therefore, in
section~\ref{sect:Xsects}, positron-proton cross-sections are quoted
for $Q^{2} < 4$~GeV$^2$. Events with higher $Q^2$ are estimated to contribute
less than 1\% to the sample.

\begin{figure}[ht]
\setlength{\unitlength}{1cm}
\begin{picture}(16.0,16.0)
\put(-1.0,0.0){\epsfig{figure=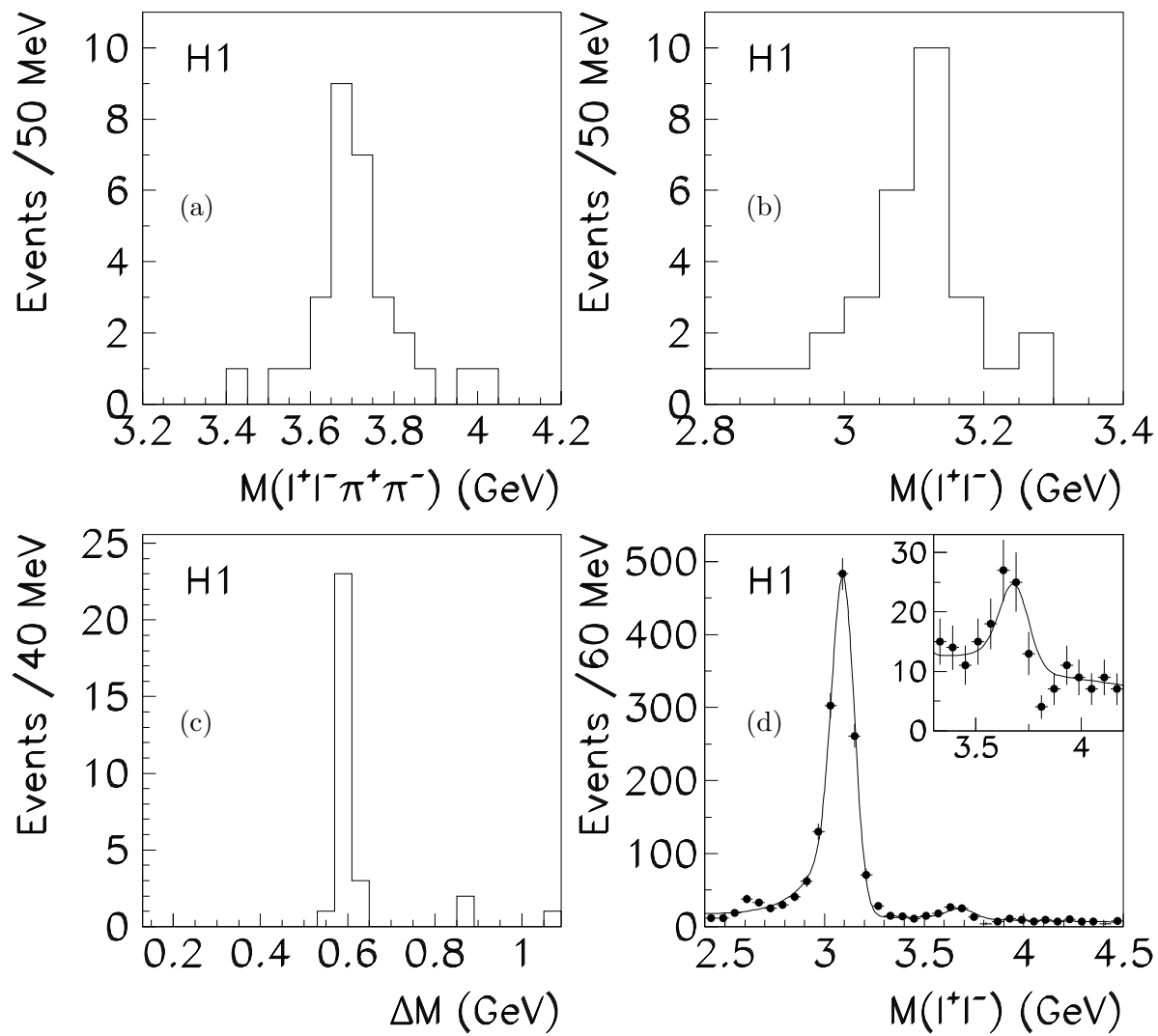,width=18.0cm}}
\put(2.2,12.5){(a)}
\put(10.0,12.5){(b)}
\put(2.2,5.4){(c)}
\put(10.0,5.4){(d)}
\end{picture}
\caption{The effective mass plots for the four track sample: 
(a) the four track effective mass,
(b) the di-lepton effective mass, and
(c) the $\Delta M$ distribution
where $\Delta M = M_{(l^{+}l^{-}\pi^{+}\pi^{-})} - M_{(l^{+}l^{-})}$.
(d) Effective mass distributions for the two track sample:
the di-lepton effective mass with a fit to two gaussians, convoluted
with an exponential tail to account for di-electron decays in which a decay 
electron has lost energy through radiation before entering the 
detector, plus a linear background.
}
\label{fig:masses}
\end{figure}

The mass distributions obtained are shown in figures~\ref{fig:masses}(a-d),
where the data from 1994 and 1995 and for both di-lepton types are
combined.
Figure~\ref{fig:masses}(c) shows a clear signal of 27 events 
with negligible background in the range 
$M_{\psip} - M_{J/\psi} - 60 < \Delta M < M_{\psip} - M_{J/\psi} + 60$~MeV 
where 
$\Delta M \equiv 
M_{(\ell^{+}\ell^{-}\pi^{+}\pi^{-})} - M_{(\ell^{+}\ell^{-})} $.
The number of events in the 
$\psip \rightarrow \ell^{+}\ell^{-}$ signal is estimated by fitting the
mass distribution to two gaussian functions, convoluted with an exponential
tail to account for di-electron decays in which one of the decay electrons
has lost energy through radiation,
 above a linear background 
(figure\ref{fig:masses}(d)). A signal of $50 \pm 12$ events is obtained.
No contribution from $\psi $(3770) to these signals is expected.


\section{Monte Carlo Model and Acceptances} \label{sect:MC}
Simulation studies were made to
obtain the acceptances and efficiencies 
of triggering, track reconstruction, selection and lepton identification 
for the $\psip$ production processes. These studies used 
event samples generated with the DIFFVM Monte Carlo generator 
\cite{ref:DIFFVM}, and a detailed simulation of the H1 detector response
based on the GEANT program~\cite{ref:GEANT}.
DIFFVM generates events according to the cross-section dependence
${\rm d}\sigma/{\rm d}t \propto \Wgp^{4\varepsilon} e^{bt}$ for elastic
$\psip$ production, and 
${\rm d}^{2}\sigma/{\rm d}t \, {\rm d}\! M_{Y}^{2} \propto \Wgp^{4\varepsilon} 
e^{b^{\prime}t} M_{Y}^{\beta}$ for production with proton dissociation,
where $M_{Y}$ is the effective mass of the hadrons produced in the 
dissociation of the proton which is related to the elasticity $z$ by
$z \approx (W^{2} - M_{Y}^2)/(W^{2} - m_{p}^2)$.
The parameters of the DIFFVM generator  
are chosen such that the features of $J/\psi$ 
elastic and proton dissociation photo-production processes are reproduced 
\cite{ref:H1_Jpsi2,ref:H1_Jpsi3}: $4\varepsilon = 0.9$,
$b = 4.0$ GeV$^{-2}$, $b^{\prime} = 1.6$ GeV$^{-2}$, $\beta = -2.16$. 
Equal cross-sections for elastic
and proton dissociative photo-production of $\psip$ are assumed, 
consistent with the results of the $J/\psi$ analysis.
The decay angular distribution
for $\psip$ decaying directly into
two leptons is simulated according to $s$-channel helicity conservation. 
The Monte Carlo program is found to model all features of the data well,
in particular the momentum and angular distributions of the $\psip$ decay
products to which the acceptance is sensitive.

The acceptance is found to be not strongly dependent on $\Wgp$ in the
range $40 < \Wgp < 120(160)$~GeV for the electron(muon) decay channel.
By demanding that the measured value of $\Wgp$ lies in this 
region the sensitivity
of the measured cross-section to the uncertainty in its $\Wgp$ 
dependence is small.
The requirement that there be no
charged particle observed in the detector, other than the decay products
of the $\psip$ meson, restricts the acceptance approximately to the
region $z>0.95$. 
Events with $z<0.95$ are estimated to contribute a background
of less than 1\% to the signal.
Within the kinematic region $z>0.95$ and $40 < \Wgp < 120(160)$~GeV
the total probability of observing a produced $\psip$ meson is about $6$\% 
for the four-particle decays and $18$\% for the two-particle
decays. 

Uncertainties in the acceptance calculations contribute to systematic
uncertainties in the cross-section measurements. The uncertainty due
to the use of the Monte Carlo model is estimated to be $5\%$ by 
observing the changes in the acceptance resulting from varying the
parameters $4\varepsilon$ by $\pm 0.1$, $b$ and $b^{\prime}$ by
$\pm 1.0$~GeV$^{-2}$, $\beta$ between $-2.00$ and $-2.50$, and by
changing the assumed fraction of the quasi-elastic cross-section 
due to elastically produced $\psip$ by $\pm 20$\%. 
Further systematic uncertainties arise due to uncertainties in the
precision of the simulation of the H1 detector response. The most
significant are $12$\% due to the track reconstruction efficiency
($6$\% for the two-particle decays), $6$\% due to the lepton identification
efficiency and $6$\% due to the trigger efficiency. Additionally
for the two-particle decays, there is a $20$\% systematic uncertainty
in the cross-section measurement resulting from the choice of the
functional form of the fit to the di-lepton mass distribution in
figure~\ref{fig:masses}(d).
The total systematic uncertainties in the cross-section measurements
are $16$\% for the four-particle decays and $23$\% for the
two-particle decays. Branching ratio (BR) uncertainties of $9$\% and
$18$\% respectively~\cite{ref:PDG} are given separately below.


\section{\boldmath{$\psip$} Cross-sections} \label{sect:Xsects}

The numbers of events, acceptances, flux factors and
cross-sections are given in table~\ref{table:results}. For the
four-particle decays of the $\psip$ the results are shown separately
for the electron and muon decays, and for the 1994 and 1995 data
samples. The corresponding four cross-sections are consistent with
one another. A decomposition of the acceptance is given, which shows
that, except for the trigger efficiencies, aspects of the acceptance
are stable from year to year and decay channel to decay channel.
The higher trigger efficiencies in 1994 with respect to 1995 reflect
a higher fraction of events in which the trigger fired being recorded
for subsequent analysis. This fraction is determined by HERA beam
background conditions and global data-taking requirements. 
For the two-particle decays all data are
combined in table~\ref{table:results} because of the lack of statistics
of the subsamples in the presence of substantial background. 

The combined cross-section from the four-particle decays using 1994 and 1995 
data is
\[
 \sigma \left[ e^{+} p \rightarrow e^{+}\psip Y \right] = 
\left[ 2.02 \pm 0.39({\rm stat})
\pm 0.33({\rm syst}) \pm 0.18({\rm BR})
\right] ~{\rm nb} 
\]
for the kinematic range $40 < W_{\gamma p} < 160$~GeV, $Q^{2} < 4$~GeV$^2$,
$z>0.95$.

The photo-production ($Q^2 = 0$~GeV$^2$) cross-section at 
$W_{\gamma p}=80$~GeV is related to the electro-production cross-section by
\begin{equation}
\sigma _{\rm ep} = \sigma _{\gamma p}(W_{\gamma p}=80~{\rm GeV}) 
\int {\rm d}y \int {\rm d}Q^{2} 
f_{\gamma/e}(y,Q^{2}) 
\left( \frac{W_{\gamma p}}{80~{\rm GeV}} \right) ^{4\varepsilon}
\left( 1 + \frac{Q^{2}}{M_{\psip}^{2}} \right)^{-n}
\label{eqn2} \end{equation}
where $W_{\gamma p} = m_p^2 - Q^2 + y(s - m_p^2 - m_e^2)$, 
$4\varepsilon = 0.9$, $n = 2$.
The limits of integration are determined by the kinematic range
of the electro-production cross-section. The spectral flux
of transverse photons from the electron~\cite{ref:budnev} is
\begin{equation}
f_{\gamma/e}(y,Q^{2})=\frac{\alpha}{2~\pi~Q^{2}} \cdot
\left[ 1 + (1-y)^{2} - \frac{2 m_{e}^{2} y^{2}}{Q^{2}} \right] . 
\end{equation}
Given an approximate $W_{\gamma p}^{0.9}$ dependence of the photo-production 
cross-section, the choice of $W_{\gamma p}=80$~GeV as the point at which to 
specify $\sigma _{\gamma p}$ minimises the uncertainty in $\sigma _{\gamma p}$
due to possible deviations from this dependence.
Varying the assumed $W_{\gamma p}$  and $Q^2$ 
dependences of the virtual photon-proton cross-section
in equation~\ref{eqn2} ($4\varepsilon $ between $0.8$ and $1.0$ and
$n$ between $1$ and $3$) 
results in a systematic 
uncertainty in the photo-production cross-section of 3\%.

For $z>0.95$ at $\Wgp =80$~GeV, the photoproduction cross-section
\[ \sigma \left[ \gamma p \rightarrow \psip Y \right] = 
\left[ 16.9 \pm 3.3({\rm stat}) \pm 2.7({\rm syst}) \pm 1.5({\rm BR})
\right] ~{\rm nb} \]
is obtained.

The $\psip$ signal of $50\pm 12$ events from the analysis
of the two-particle decay sample yields a statistically independent 
measurement of
the cross-sections for $\psip$ production:
\[ 
\sigma \left[ e^{+}p \rightarrow e \psip Y \right] = 
\left[ 2.65 \pm 0.63({\rm stat}) \pm 0.61({\rm syst}) \pm 0.48({\rm BR})
\right] ~{\rm nb,}
\]
\[
\sigma \left[ \gamma p \rightarrow \psip Y \right] = 
\left[ 22.1 \pm 5.3({\rm stat}) \pm 5.1({\rm syst}) \pm 4.0({\rm BR})
\right] ~{\rm nb}
\] 
for the kinematic intervals above.
Here the estimate of the level of background under the $\psip$ signal
contributes a $20$\% systematic uncertainty.

The results from the samples of two-particle and four-particle decays 
agree well, so they
are combined to give an overall photo-production cross-section of
\[
\sigma \left[ \gamma p \rightarrow \psip Y \right]  = 
\left[ 17.9 \pm 2.8({\rm stat}) \pm 2.7({\rm syst}) \pm 1.4({\rm BR})
\right] ~{\rm nb}
\]
at $\Wgp = 80$~GeV, for $z>0.95$. 


\section{The Cross-section Ratio 
\boldmath{$\sigma \left[ \psip \right]$/$\sigma \left[ \Jpsi \right]$} } 
\label{sect:ratio}

\begin{figure}[ht]
\begin{center}
\epsfig{figure=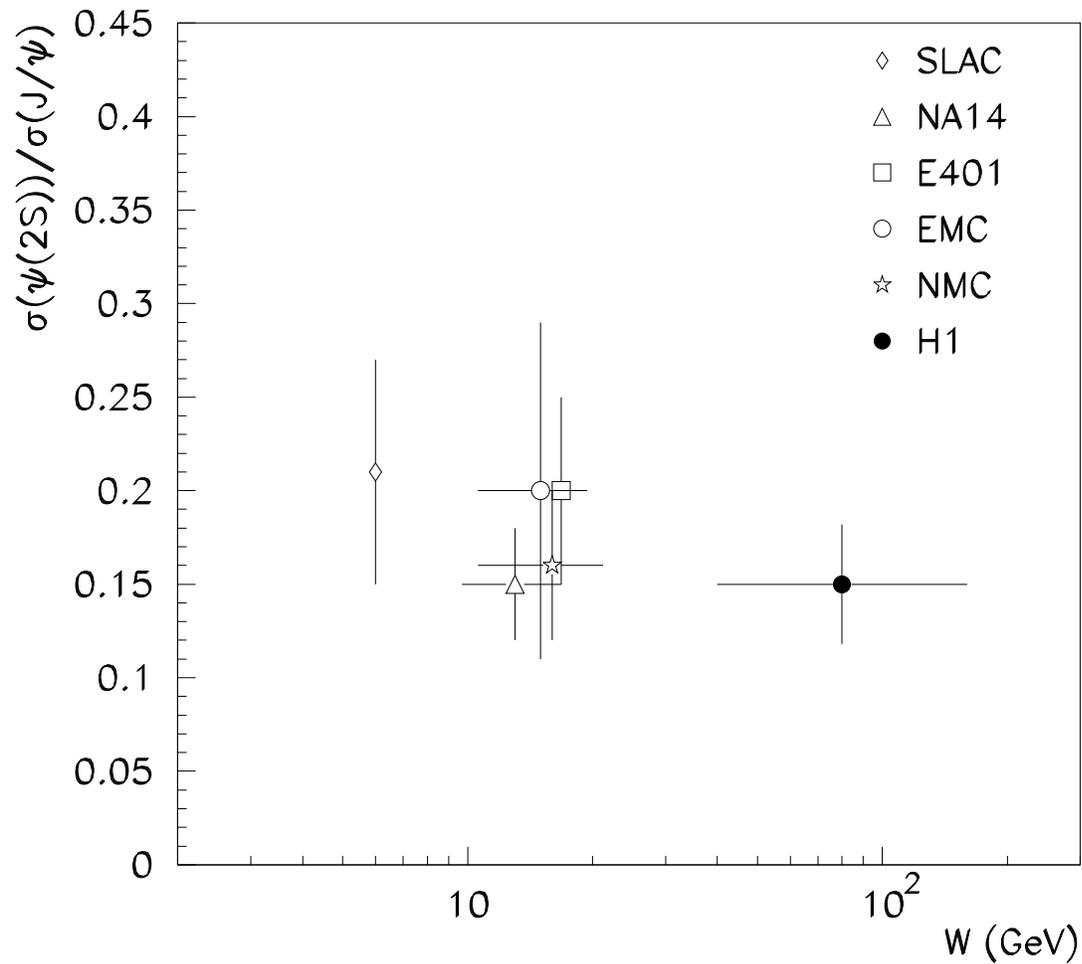,width=16.0cm}
\end{center}
\caption{The ratio of the cross-sections for $\psip$ and $\Jpsi$
photo-production on nucleons as a function of photon-proton 
centre-of-mass energy,
$\Wgp$. The results from the present analysis are shown together with
those from earlier fixed target experiments 
\cite{ref:SLAC_prime,ref:NA14_prime,ref:E401_prime,ref:EMC_prime,ref:NMC_prime}. The ratios from the fixed target experiments 
have been corrected using the latest decay branching ratios. The
error bars shown on the low energy measurements are statistical only.
The combined statistical and systematic errors are shown for the
H1 result.
Horizontal bars indicate the ranges of $\Wgp$ over which the 
measurements were made, where they are specified. The common uncertainty
due to the imprecisely known branching ratios is not shown.
}
\label{fig:comp_new}
\end{figure}

The $J/\psi$ quasi-elastic ($z>0.95$) photo-production cross-section 
at $\Wgp =80$~GeV is
$(119.5\pm11.2\pm12.0)$ nb \cite{ref:H1_Jpsi3}, yielding a ratio:
\[ 
\frac {\sigma \left[ \gamma p \rightarrow \psip Y \right]}
{\sigma \left[ \gamma p \rightarrow J/\psi Y \right] } = 
0.150 \pm 0.027({\rm stat}) \pm 0.018({\rm syst}) \pm 0.011({\rm BR}).
\] 
Figure \ref{fig:comp_new} contains a compilation of measurements
from fixed target experiments of the ratio of the photon-nucleon cross-sections
for the production of $\psip$ and $J/\psi$ mesons:
three
photo-production experiments at $Q^{2}=0$~GeV$^2$ (deuterium target 
\cite{ref:SLAC_prime} SLAC, 
lithium target \cite{ref:NA14_prime} NA14, and deuterium 
target \cite{ref:E401_prime} E401) plus 
muo-production
experiments on an iron target \cite{ref:EMC_prime}(EMC) and more
recently on tin and carbon \cite{ref:NMC_prime}(NMC). Of these the deuterium 
target data samples were restricted to events in which the decay products of
the vector mesons were the only tracks detected; in the lithium 
and iron target
data some contribution from the kinematic region of lower $z$ was included.
Despite the different experimental conditions, it is apparent
that in going from fixed target to HERA energies there is no significant 
change in the $\psip$ to $J/\psi$ cross-section ratio. 


\section{Colour Dipole Calculation} \label{sect:cdcalc}

By considering the $c \bar{c}$ quark pair as a colour dipole of
transverse size $r$, cross-sections for the photo-production of
charmonium can be calculated in the framework of 
QCD~\cite{ref:Kop91,ref:Kop93}. The amplitude of forward photo-production
on the proton can be written
\begin{equation}
\int d^2 \vec{r} \, \, \,
\Psi _V (\vec{r})^{*} \sigma (\vec{r}) \Psi _{\gamma} (\vec{r})
\end{equation}
where $\Psi _{\gamma}(\vec{r})$ is
the wave-function representing the probability of finding a $c \bar{c}$
dipole of size $r = \left| \vec{r} \right|$ in the photon, 
$\sigma (\vec{r})$ is the colour dipole
cross-section, and $\Psi _V (\vec{r})$ the transverse wave-function of the
charmonium state. 
The product $\sigma (\vec{r}) \Psi _{\gamma} (\vec{r})$ is calculated
in QCD using the approximation of a 
two-gluon exchange between the colour dipole and the proton.
It is found that, approximately, $\Psi _{\gamma} (\vec{r}) \propto 
{\rm exp}(-m_{c}r)$ and $\sigma (\vec{r}) \sim r^2$, so that the product is
a function peaking at $r \approx 2/m_{c}$. As a result the matrix element,
and hence the cross-section, is sensitive to the radial wave-function of
the charmonium around $r = 2/m_{c}$.
This feature influences the ratio of the cross-sections for the
production of $\psip$ and $J/ \psi (1S)$ mesons. 
There is a relative suppression of the $\psip$ because its wave-function
has a radial node (the value of $r$ for which $\Psi _V (\vec{r}) = 0$)
near to $r = 2/m_{c}$, in contrast to the $J/ \psi$ wave-function. 
Therefore the ratio of these cross-sections provides a test of the model
which is free of many normalisation uncertainties.

A prediction for the ratio of the forward 
elastic photo-production
cross-sections of 0.17 is obtained~\cite{ref:Kop93}, 
which agrees well with the
ratio of quasi-elastic cross-sections reported here.
This agreement supports the hypothesis that the charmonium production
cross-section is sensitive to the radial wave-function of the
meson away from the origin, resulting in a suppression of the
production of the $2S$ state, $\psip$, with respect to the
$1S$ state, $\Jpsi$.

\section{Summary}
Measurements of the quasi-elastic photo-production
cross-section for $\psip$ at a $\Wgp$ value of 80 GeV at HERA
have been reported. The ratio between the quasi-elastic 
$\psip$ and $J/\psi$ photo-production
cross-sections is compatible with those measured at lower
energies.
A prediction from a QCD based model in which the photon fluctuates into
a $c \bar{c}$ quark pair which then behaves as a colour dipole
in its interaction with the proton is in good agreement with the measured 
ratio. 

\section*{Acknowledgments} 
We are grateful to the HERA machine group whose outstanding
efforts have made and continue to make this experiment possible. We thank
the engineers and technicians for their work in constructing and 
maintaining the H1 detector, our funding agencies for financial support, the
DESY technical staff for continual assistance, and the DESY directorate for the
hospitality which they extend to the non--DESY members of the collaboration.

\begin{table}[p]
\begin{tabular}{|c|c|c|}
  \hline
  Data  & $e^+e^-\pi^+\pi^-$ & $e^+e^-\pi^+\pi^-$  \\
  sample &  (1994)   &    (1995)   \\
  \hline
  $W_{\gamma p}$ (GeV) & $40-120$ & $40-120$ \\
  \hline
  $Q^2$ (GeV$^2$)    & $Q^2<4.0$ & $Q^2<1.2$ \\
  \hline
  Luminosity (pb$^{-1}$) & 2.70 & 3.59  \\
  \hline
  Angular acceptance & 0.495 & 0.492  \\
  \hline
  $p_t$ track acceptance & 0.715 & 0.704  \\
  \hline
  Trigger efficiency     & 0.431 & 0.195  \\
  \hline
  Analysis efficiency    & 0.514 & 0.514  \\
  \hline
  Total acceptance       & 0.078    & 0.035  \\
  \hline
  Signal events      & 5 & 4 \\
  \hline
   \boldmath{$\sigma(ep \rightarrow e \psip Y)$}{\bf /nb} &
  \boldmath{$1.21 \pm .54 \pm .20$} & 
  \boldmath{$1.64 \pm .82 \pm .33$}   \\
  \hline
  $\Phi_{\gamma/e}$  &  0.088  &  0.085  \\
  \hline
  \hline
  \boldmath{$\sigma(\gamma p \rightarrow \psip Y)$}{\bf /nb} &
  \boldmath{$13.7 \pm 6.1 \pm 2.3$} &  
  \boldmath{$19.3 \pm 9.6 \pm 3.9$}                  \\
  \hline
 \end{tabular}

\vspace{0.5cm}
\begin{tabular}{|c|c|c|c|}
  \hline
  Data  & $\mu^+\mu^-\pi^+\pi^-$ & $\mu^+\mu^-\pi^+\pi^-$ & 
  $\ell^+\ell^- (\ell=\mu {\rm ,} e)$ \\
  sample &    (1994)      &   (1995) & (1994+1995)  \\ \hline
  $W_{\gamma p}$ (GeV) & $40-160$ & $40-160$ & $40-160$\\ \hline
  $Q^2$ (GeV$^2$)    & $Q^2<4.0$ & $Q^2<1.2$ & $Q^2<4.0$  \\ \hline
  Luminosity (pb$^{-1}$) & 2.72 & 3.63 & 6.37      \\ \hline
  Angular acceptance & 0.628 & 0.622  & 0.614     \\ \hline
  $p_t$ track acceptance & 0.715 & 0.712 & 0.994  \\ \hline
  Trigger efficiency     & 0.367 & 0.207  &  0.487   \\ \hline
  Analysis efficiency    & 0.518 & 0.606 & 0.588    \\ \hline
  Total acceptance       & 0.085 & 0.055 & 0.178    \\ \hline
  Signal events      & 10 & 8 & 49.6 \\ \hline
  \boldmath{$\sigma(ep \rightarrow e \psip Y)$}{\bf /nb} 
  & \boldmath{$2.21 \pm .70 \pm .37$} &  \boldmath{$2.06 \pm .73 \pm .39$}  & 
  \boldmath{$2.65 \pm .63 \pm .61$}                           \\ \hline
  $\Phi_{\gamma/e}$  &  0.120  &  0.115  & 0.120 \\ \hline
  \hline
  \boldmath{$\sigma(\gamma p \rightarrow \psip Y)$}{\bf /nb} &
  \boldmath{$18.5 \pm 5.8 \pm 3.1$} & \boldmath{$17.9 \pm 6.3 \pm 3.4$} & 
  \boldmath{$22.1 \pm 5.3 \pm 5.1$}                          \\ \hline
 \end{tabular}
 \caption
 {Summary of signals, acceptances, and the quasi-elastic ($z>0.95$) 
 electro- and photo-production cross-sections for the five 
 data samples. The first error shown is statistical and the second 
 systematic. Uncertainties due to the imprecisely known branching
ratios of the $\psip$ decays are not included. 
The ranges over $\Wgp$ and $Q^2$ over which the $ep$ cross 
 section measurement is made are shown. Angular and $p_t$ track
 acceptance refer to the fraction of events in the kinematic range in
 which all decay products fall within the angular and $p_t$ ranges
 of acceptance, respectively. Trigger efficiency indicates the probability
that such an event is accepted by the first level trigger, and the analysis
efficiency gives the probability of subsequently meeting all further selection
criteria (see section~\ref{sect:ident}). $\Phi_{\gamma/e}$ is the photon 
 flux factor used to give photo-production 
 cross-sections at a fixed $W_{\gamma p}$ of $80$~GeV.}
 \label{table:results}
\end{table}

\newpage

\newpage

\end{document}